\documentclass[twocolumn,showpacs,preprintnumbers,amsmath,amssymb]{revtex4}

\usepackage{graphicx}
\usepackage{dcolumn}
\usepackage{bm}

\begin{document}

\title{Emerging Zero Modes for Graphene in a Periodic Potential}

\author{L. Brey}
\affiliation{Instituto de Ciencia de Materiales de Madrid (CSIC),
Cantoblanco 28049, Spain}
\author{H.A. Fertig}
\affiliation{Department of Physics, Indiana University, Bloomington, IN 47405}

\date{\today}

\begin{abstract}
We investigate the effect of a periodic potential on the electronic
states and conductance of graphene.  It is demonstrated that for a
cosine potential $V(x)=V_0\cos(G_0x)$, new zero energy states emerge
whenever $J_0(\frac {2V_0}{\hbar v_F G_0})=0$.  The phase of the
wavefunctions of these states is shown to be related to periodic
solutions of the equation of motion of an overdamped particle in a
periodic potential, subject to a periodic force.  Numerical
solutions of the Dirac equation confirm the existence of these
states, and demonstrate the chirality of states in their vicinity.
Conductance resonances are shown to accompany the emergence of these
induced Dirac points.
\end{abstract}
 \maketitle

The laboratory realization of graphene
\cite{Novoselov_2004,Novoselov_2005,Zhang_2005}, a two-dimensional
honeycomb lattice of carbon atoms, has motivated intense theoretical
and experimental investigations of, among many other things, its
transport properties \cite{Castro_Neto_RMP}.  Graphene differs from
conventional two-dimensional electron systems in that it supports
two inequivalent Dirac points, through which the Fermi energy passes
when the system is nominally undoped.  The chirality of the electron
wavefunctions near the Dirac points severely limits electron
backscattering, enhancing the conductivity of the system
\cite{Ando_Nakanishi_1998}. The unusual transport properties
associated with the Dirac points may in principle be explored and
exploited if the edge structure can be controlled
\cite{Brey_2006b,Iyengar_2008}, or by application of non-uniform
potentials, such as in a {\it pn} junction
\cite{Huard_2007,Ozyilmaz_2007,Williams_2007,Young_2009,Stander_2009}.
Periodic potentials may be induced by interaction with a substrate
\cite{Marchini_2007,Parga_2008,Pan_2009} or controlled adatom
deposition \cite{Meyer_2008}. Recently, the existence of periodic
ripples in suspended graphene has been demonstrated \cite{Lau_KITP};
in a perpendicular electric field this would also induce a periodic
potential.

In this paper we discuss the effects of a one-dimensional
superlattice potential on the transport properties of graphene.
Previously it has been noted that such a potential may create a
strong anisotropy in the electron velocity around the Dirac point
\cite{Park_2008a}. In fact this behavior is a precursor to the
formation of further Dirac points in the band-structure of the
system. We will show that such new zero energy states of the Dirac
equation are associated with how the phase of the wavefunction
varies with position along the superlattice direction, and is
connected with the dynamics of a highly overdamped particle subject
to periodic potentials both in space and time.  From an analysis of
this problem we demonstrate that the emergence of new Dirac points
is controlled by the parameter $V_0/G_0$, where $V_0$ is the
potential amplitude (assumed to be a cosine) and $L_0=2\pi/G_0$ is
the period.  New Dirac points emerge whenever $J _0 ( \frac {2
V_0}{\hbar v_F G_0})=0$, where $J_0$ is a Bessel function and $v_F$
the speed of the Dirac fermions in the absence of the potential. The
total number of Dirac points (associated with a single valley and
electron spin) is thus $2N+1$, with $N$ the number of zeros of
$J_0(x)$, with $|x|<\frac {2 V_0}{\hbar v_F G_0}$.  We verify the
existence of these new zero energy points numerically, and
demonstrate the chirality of the wavefunctions in their vicinity.

\begin{figure}
  \includegraphics[clip,width=9cm]{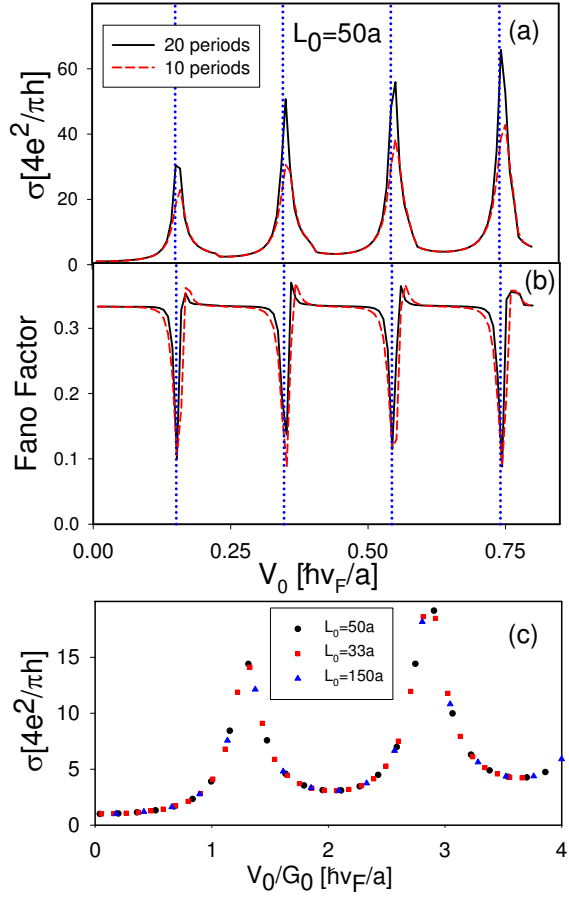}
  \caption{($Color$ $online$)(a) Conductivity and (b) Fano factor,
 as function of $V_0$ and for $L_0=50a$,
  for two different graphene
  sample lengths containing
  10 and 20 periods of the periodic potential $ V_0 \cos G_0 x$.
  Panel (c) shows the conductivity as function of $V_0/G_0$ as
  obtained from different superlattices potentials. This result
  indicates that the conductivity only depends on $V_0/G_0$. Vertical ines indicates the
  position of the zeros of $J _0 ( \frac {2
V_0}{\hbar v_F G_0})$}
   \label{Figure1}
\end{figure}

The chiral nature of the wavefunctions around the graphene Dirac
points in the absence a periodic potential has dramatic consequences
for its transport properties. We find that, when stabilized, the new
Dirac points associated with the superlattice potential have
important consequences as well, for transport along the superlattice
axis. For undoped graphene, strong conductance resonances appear at
the values of $V_0/G_0$  where they first appear, as illustrated in
Fig. \ref{Figure1}.  Interestingly, in the limit of large width, the
conductance scales for most values of $V_0/G_0$ as $L_y/L_x$, with
$L_y$ the system width and $L_x$ its length, indicating the system
behaves diffusively.  This interpretation is supported by the Fano
factor, as illustrated in Fig. \ref{Figure1}(b). At the resonances,
however, the conductance becomes independent of $L_x$, and the Fano
factor indicates a more ballistic-like transport. The periodic
appearance of conductance resonances as a function of $V_0/G_0$
provides a clear signature of new zero energy states as they emerge.

\textit{Zero Energy States} --  We consider an external potential of
the form $V(x)= V_0 \cos{G_0 x}$,  where the period of the
perturbation is much larger than the graphene lattice parameter,
$a$, and the amplitude $V_0$ is much smaller than the energy
bandwidth of the graphene $\pi$-orbitals. In this situation the low
energy properties for electrons in a single valley and a given spin
are well described by the massless Dirac Hamiltonian with a
potential,
\begin{equation}
H = \hbar v_F (-i \sigma _x \partial _x - i \sigma _ y \partial
_y) +V(x){\cal I}\label{Dirac}
\end{equation}
where $\sigma_{x,y}$ are the Pauli
matrices, and ${\cal I}$ is the identity matrix.
The wavefunctions which this may multiply have two components,
$\Phi_{A,B}$, corresponding
to the
two triangular sublattices that make up a honeycomb lattice.

To demonstrate the emergence of zero energy states, it is convenient to implement
a unitary  transformation\cite{Talyanskii_2001,Park_2008c}
of the Hamiltonian, $H ^ {'} = U_1 ^{\dag} H U _1 $, with
\begin{equation} U_ 1 \! = \!
\left(
  \begin{array}{cc}
    e ^{-i \frac {\alpha (x)} 2}  & -e ^{i \frac {\alpha (x)} 2} \\
    e ^{-i \frac {\alpha (x) }2} & e ^{i \frac {\alpha (x)}2}  \\
  \end{array}
\right) \textrm{and } \alpha (x) \! = \! \frac 2 {\hbar v _F} \!
\int _0 ^x \! \!V(x') d x',\label{Unitary}
\end{equation}
so that, for $\Phi_{A,B}({\bf r})=e^{ik_yy}\phi_{A,B}(x)$,
the Hamiltonian acting on the transformed wavefunctions $U^{\dag}\vec{\phi}$ is
\begin{equation}
H^{\prime} =  \hbar v_F \left(
  \begin{array}{cc}
    -i \partial _x & -i k_y e ^{i \alpha (x)} \\
     i k_y e ^{-i \alpha (x)} & i \partial _x \\
  \end{array}
\right). \label{hprime1}
\end{equation}
For the cosine potential, $e ^{i \alpha (x)}= \sum _{l= - \infty}
^{l=\infty} J _{\ell} ( \frac {2 V_0}{\hbar v_F G_0}) e ^{i l G_0 x}$,
where $J_n$ is the $n$-th Bessel function of the first kind.

For zero energy states, we search for solutions of $H^{\prime}\vec{\phi}=0$. Such
solutions have the property $\phi_A=\phi_B^*$.  For a single valley system
this would mean the two components are related by time-reversal;
this is not quite the case here because time reversal in graphene also
involves interchanging valleys.  Writing $\phi_A=|\phi_A|e^{i\chi}$,
we obtain a single component equation
\begin{equation}
\partial_x|\phi_A|+k_y e^{i\alpha-2i\chi}|\phi_A| + i (\partial\chi)|\phi_A|=0.
\label{phi_eq1}
\end{equation}
An analogous manipulation for the zero modes of the Bogoliubov-de
Gennes equation of states in a vortex of a p-wave superconductor
\cite{Read_2000} yields a very simple value for the phase $\chi$,
and the resulting state is a Majorana fermion due to the
time-reversal relation between components. In our case, we need to
find solutions for $\chi$ that are consistent with the symmetries of
the problem; when they exist the resulting state is that of a real
fermion, since the two magnitudes $\phi_{A,B}$ are not truly related
by time-reversal. The equation governing $\chi$ is obtained by
taking the imaginary part of Eq. \ref{phi_eq1},
\begin{equation}
k_y\sin(\alpha-2\chi)+\partial_x\chi=0.
\label{chi_eq}
\end{equation}
The real part yields
$
\partial_x|\phi_A|+k_y\cos(\alpha-2\chi)|\phi_A|=0,
$
with the formal solution
\begin{equation}
|\phi_A| \propto \exp{ \left\{-k_y\int_{x_0}^{x}
\cos[\alpha(x^{\prime})-2\chi(x^{\prime})]dx^{\prime} \right\}}.
\label{phi_eq2}
\end{equation}

Since $\vec{\phi}$ is a Bloch state of the superlattice, it must obey the Bloch relation
$\phi_{A,B}(x+L_0)=e^{ik_xL_0}\phi_{A,B}(x)$,, with $k_x$ the crystal momentum.
For a zero energy state only $k_x=0$ is possible.  We then require
({\it i}) $\chi(x+L_0)=\chi(x)+2\pi m$ with $m$ an integer, and ({\it ii}) $\int_0^{L_0}\cos[\alpha(x)-2\chi(x)]=0$.
To see whether $\chi$ can satisfy these relations, it is helpful
to recast Eq. \ref{chi_eq} by writing $\tilde{\chi}=2\chi-\alpha$, and $x \rightarrow t$, so
that
\begin{equation}
-\partial_t \tilde{\chi} -\partial_t{\alpha}+2k_y\sin \tilde{\chi}=0.
\label{swimmer}
\end{equation}
This is the equation of motion for the position $\tilde{\chi}$
of an overdamped particle (with unit viscosity), subject to
a periodic time-dependent force $\partial_t{\alpha}$ and a spatially periodic force
$2k_y\sin \tilde{\chi}$.  Despite the periodicity of the forces involved,
the generic solution to this equation is not periodic.  However, for certain
parameters periodic solutions can be found, which correspond to allowed zero
energy solutions of the Dirac equation in a periodic potential.

To see this, we solve Eq. \ref{chi_eq} perturbatively in $k_y$.  Writing
$\chi=k_y\chi^{(1)}+k_y^2\chi^{(2)}+{\cal O}(k_y^3)$, one finds
\begin{eqnarray}
\chi^{(1)} \! \!& = &\! \! -\int^{x}dx^{\prime} \sin
\alpha(x^{\prime})+C^{(1)}
\, \, \textrm{and} \nonumber \\
\chi^{(2)}\! \! & = & \! \! 2C^{(1)} \! \! \int^{x} \! \! \! dx_1
\cos\alpha(x_1) \! - \! \! \int^x \! \! \! dx_1 \! \! \! \int^{x_1}
\! \! \! \!  dx_2 \sin \alpha(x_2) \! + \! C^{(2)}, \nonumber
\end{eqnarray}
where $C^{(1,2)}$ are constants of integration.  Explicitly performing the
integrations for the above two equations, one finds that condition ({\it i}) can be satisfied
if
\begin{equation}
k_y^2\left[2 C^{(1)}J_0L_0 - \sum_{\ell \,odd} \frac{J_{\ell}}{\ell G_0} L_0 \right] = 2\pi m.
\label{c(1)_eq}
\end{equation}
Here $J_0$ and $J_{\ell}$ are Bessel functions evaluated at
$2V_0/\hbar v_F G_0$.  Since we have employed a small $k_y$
expansion, the only consistent solution is for $m=0$. In this case
Eq. \ref{c(1)_eq} fixes $C^{(1)}$, and the resulting $\chi$ (and the
associated $\tilde{\chi}$) is periodic.  Condition ({\it ii}) may
then be implemented to fix  the value of $k_y$ at which a zero mode
appears,
\begin{equation}
\left(\frac{k_y}{G_0}\right)^2=-\frac{J_0}{2\sum_{\ell_1,\ell_2\,odd}
J_{\ell_1} J_{\ell_2}J_{-\ell_1-\ell_2}/\ell_1\ell_2}.
\label{ky_eq}
\end{equation}
Eq. \ref{ky_eq} predicts the presence of a zero mode whenever the
right hand side is positive.  This turns out to occur when
$x = \frac {2 V_0}{\hbar v_F G_0}$ is just above the values for which
$J_0(x)=0$; we have confirmed numerically that the sign of the
denominator on the right hand side of Eq. \ref{ky_eq} always works
out such that $k_y^2>0$ in this situation.  With increasing $x$, the
solution moves to larger $|k_y|$ until it diverges where the three
Bessel function sum vanishes, which is always prior the next zero of
$J_0(x)$.  We note that since our approximation is only valid for
small $k_y$, Eq. \ref{ky_eq} cannot accurately predict the location
of the zero energy states well away from $k_y=0$. However, since
zero energy states can only annihilate in pairs \cite{Gurarie_2007},
we expect that once they emerge from the origin, they should
persist. We now show that a numerical solution of the Dirac equation
supports this expectation.

\textit{Numerical Studies} -- Our expectations about the new zero
energy states can be directly confirmed by numerically solving the
Dirac equation in a periodic potential. To accomplish this we
represent the Hamiltonian $H=H_0+ V_0 \cos G_0 x$ in a plane wave
basis and diagonalize the resulting matrix for momenta $(k_x,k_y)$,
with $-G_0/2<k_x<G_0/2$. We have checked that our results converge
with respect to the number of plane wave states used.

\begin{figure}
  \includegraphics[clip,width=9cm]{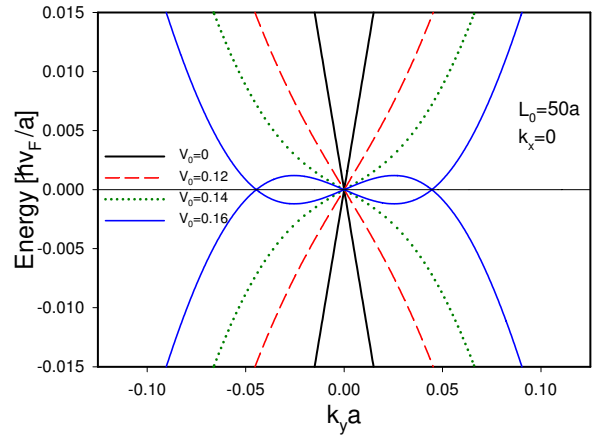}
  \caption{($Color$ $online$)Energy bands of graphene in presence
  of a superlattice potential $V_0 \cos{G_0 x}$, as function of $k_y$ with $k_x$=0, for several values of
  $V_0$, in units $\hbar v_F /a$, and $L_0=50a$.}
   \label{Figure2}
\end{figure}

Because of  the chiral nature of the Dirac  quasiparticles  the
effect of the periodic potential is highly anisotropic: at the
$k_x=0,k_y=0$ Dirac point the group velocity is unchanged in the
direction of the superlattice but is strongly reduced, even to zero
in some cases, in the perpendicular direction
\cite{Park_2008a,Park_2008b,Ho_2009}. In Fig.\ref{Figure2} we plot,
for different values of $V_0$, the lowest few energy eigenvalues as
a function of $k_y$ for $k_x$=0 and $L_0=50a$. As $V_0$ increases
the group velocity at the Dirac point decreases to zero
\cite{Park_2008a,Park_2008b,Ho_2009}, and thereafter two zero energy
states emerge from $k_y$=0 as the group velocity of the $k_y$=0
Dirac point becomes finite again. These are the new zero energy
states discussed above; we find that they emerge precisely when
$J_0(\frac {2 V_0}{\hbar v_F G_0})=0$. Upon further increase of
$V_0$, the group velocity along $k_x$ at $k_y$=0 becomes zero again
and a new pair of zero energy states emerge from $k_y$=0, again
precisely at the next zero of $J_0(\frac {2 V_0}{\hbar v_F G_0})=0$.
This pattern continues to repeat itself with increasing $V_0$.
Further studies for different periodicities confirms the prediction
that the emergence of these points depends only on the ratio
$V_0/G_0$, precisely as predicted in the analysis of the previous
section.

These zero energy points in fact represent new Dirac points, as can
be demonstrated by studying the chirality of the wavefunctions in
their vicinity. In Fig.\ref{Figure4} we plot the expectation value
of the pseudospin $(<\sigma_x>,<\sigma_y>)$ for the lowest posotive
energy eigenvalue as a function of crystal momentum. One may see
that this vector undergoes a 2$\pi$ rotation for any path enclosing
one of the zero energy states. The non-vanishing winding of this
vector for such paths is a clear signal of the Dirac-like nature of
the spectrum in the vicinity of a zero energy state.

\begin{figure}
  \includegraphics[clip,width=9cm]{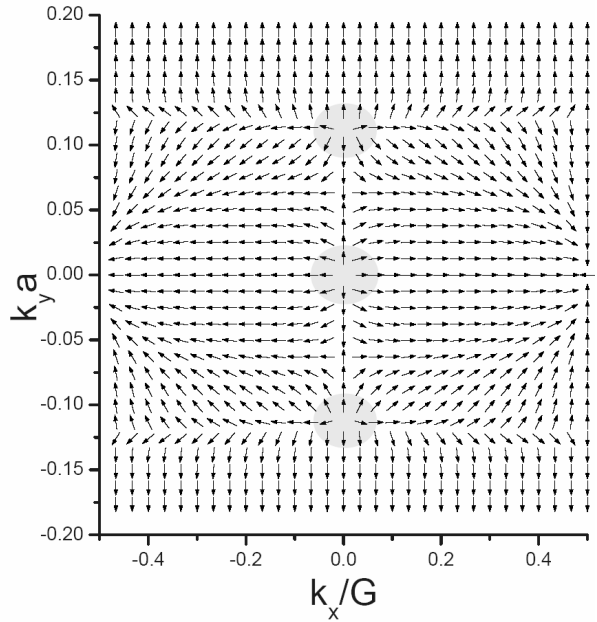}
  \caption{($Color$ $online$) Expectation value
  of the vector field $(\sigma_x,\sigma _y)$ of the lowest energy positive subband for a superlattice
  of period 50$a$ and $V_0=0.1\hbar v_F/a$. The zero energy Dirac points at $k_x$=0 are highlighted.
 High energy Dirac points with
 with opposite chirality also appear between the $k_y$=0 Dirac point and the superlattice induce Dirac points.
 }
   \label{Figure3}
\end{figure}

\emph{Conductivity}. Using the transfer matrix method we have
computed the transmission probability through a graphene strip of
length $L_x$, containing $N_p$ periods of the superlattice
potential. We assume metallic contacts connected to the strip may be
modeled by heavily doped graphene\cite{Tworzydlo_2006}. Boundary
conditions are taken to be periodic in the transverse direction,
leading to transverse wavefunctions which can be labeled by a
momentum $k_y$; this is justified when the width of the strip,
$L_y$, is much larger than its length. From the transmission
probability of each mode, $T_{k_y}$, we obtain the conductance $G$
and the Fano factor (ratio of noise power and mean current) $F$,
\begin{equation}
G=4 \frac {e ^2}{h} \sum _{k_y} T_{k_y} \, \, \, F= \frac {\sum
_{k_y} T_{k_y}(1-T_{k_y})}{\sum _{k_y} T_{k_y}} \label{conductance}
\end{equation}
where the factor 4 accounts for the spin and valley degeneracy. The
conductivity is related to the conductance via geometrical
factors, $\sigma = G \times L_y /L_x$.  In what follows we work in the limit $L_y >>L_x$.

For pristine graphene, $V_0=0$, the conductivity is independent of
$L_x$ and takes the value $\sigma _0 = 4 e^2 /\pi h$, and the Fano
factor takes on  the universal value 1/3. In the absence of disorder
one might expect the conductivity of graphene to be zero or infinity
and the the electrical current to be noiseless.  The deviation from
these naive expectations, consistent with diffusive transport, is a
unique property of the Dirac points, which has been interpreted in
terms of the spontaneous creation of virtual electron-hole
pairs\cite{Tworzydlo_2006}.

Fig.\ref{Figure1} shows the conductivity and the Fano factor, as a
function of $V_0$, for two graphene strips, respectively containing
10 and 20 periods of a potential of the form $V_0 \cos G_0 x$. For
finite values of $V_0$, apart from some resonances which we discuss
momentarily, the system behaves diffusively ($F=1/3$) and the
conductivity is well defined. Interestingly, between the peaks the
overall scale increases with $V_0$, showing that the periodic
potential tends to enhance the conductivity.

At certain values of $V_0$ one observes peaks in the
conductance, for which the conductivity is
not well defined and the Fano factor tends to zero. These resonances
occur precisely whenever new zero energy states emerge from the origin
in $k-$space, and represent a direct experimental signature of their presence.
We believe the resonances occur because the group velocity vanishes when a
zero energy state emerges, leading to a strong enhancement of the density of states.
A further check that the resonances are associated with the zero energy
states is to see that they depend on the ratio $V_0/G_0$; Fig. \ref{Figure1}(c)
demonstrates that this is the case not just for the resonances but for the
entire conductance curve.

In summary, we have shown that a search for zero energy states in graphene
in a periodic potential shows that they may be stabilized for large
enough $V_0/G_0$, and that their presence may be detected in the
conductance of the system.  The generation of new
Dirac points was very recently reported in Ref. \cite{Park_2009}, which focuses
on different properties than the ones described above.

The authors thank the KITP-UCSB where this work was
initiated.  This work was financially supported by MEC-Spain
MAT2006-03741 and by the NSF through Grant No. DMR0704033.

%

\end{document}